\documentclass[reqno,12pt]{amsart}
\usepackage{fullpage}
\usepackage{amsfonts}
\usepackage{amssymb}

\usepackage{tikz}
\usetikzlibrary{arrows.meta}
\usetikzlibrary{decorations.pathmorphing}

\numberwithin{equation}{section}
\def\alg#1{\mathfrak{#1}}

\def\posvec{{\vec r}}
\def\velvec{{\vec v}}

\def\unitposvec{{\hat r}}
\def\r{r}

\def\L{\vec{L}}
\def\A{\vec{A}}
\def\M{\vec{M}}
\def\T{\mathcal{T}}

\def\Dt{\mathcal{D}_t}

\def\Lagr{\mathcal{L}}
\def\k{\kappa}

\def\X{{\bf X}}
\def\Y{{\bf Y}}
\def\pr{{\rm pr}}

\def\P{{\vec P}}

\def\Rnum{\mathbb{R}}

\def\sgn{{\rm sgn}}

\def\arctanh{{\rm arctanh}}

\def\parder#1#2{\frac{\partial{#1}}{\partial{#2}}}
\def\totder#1#2{\frac{d{#1}}{d{#2}}}

\def\com/{constant of motion}
\def\coms/{constants of motion}
\def\eom/{equations of motion}
\def\LRL/{Laplace--Runge--Lenz}

\newtheorem{prop}{Proposition}
\newtheorem{thm}{Theorem}

\def\Ref#1{Ref.\cite{#1}}

\def\ie/{i.e.}
\def\eg/{e.g.}

\allowdisplaybreaks[3]

\begin{document}

\title{Symmetry transformation group\\ arising from the Laplace--Runge--Lenz vector}

\author{
Stephen C. Anco
\lowercase{\scshape{and}}
Mahdieh Gol Bashmani Moghadam
\\
\\\lowercase{\scshape{
Department of Mathematics and Statistics, 
Brock University\\
St. Catharines, ON, Canada}} \\
}

\begin{abstract}
The Kepler problem in classical mechanics exhibits a rich structure of conserved quantities, 
highlighted by the Laplace--Runge--Lenz (LRL) vector.
Through Noether's theorem in reverse,
the LRL vector gives rise to a corresponding infinitesimal dynamical symmetry
on the kinematical variables, which is well known in the literature. 
However, the physically relevant part of the LRL vector is its direction angle
in the plane of motion
(since its magnitude is just a function of energy and angular momentum).
The present work derives the infinitesimal dynamical symmetry
corresponding to the direction part of the LRL vector 
and obtains the explicit form of the symmetry transformations that it generates.
When combined with the rotation symmetries,
the resulting symmetry group is shown to be the semi-direct product of $SO(3)$ and $\Rnum^3$.
This stands in contrast to the $SO(4)$ symmetry group
generated by the LRL symmetries and the rotations. 
As a by-product, the action of the new infinitesimal symmetries on all of the conserved quantities is obtained.
The results are given in terms of the physical kinematical variables in the Kepler problem,
rather than in an enlarged auxiliary space in which the LRL symmetries are usually stated.
\end{abstract}

\maketitle

\section{Introduction}\label{sec:intro}
\label{intro}

The Kepler problem in Newtonian mechanics has been widely studied
and encompasses both a particle moving under inverse-square force 
and two bodies moving under mutual gravitational attraction. 
One of its most interesting features is
the existence of the conserved Laplace--Runge--Lenz (LRL) vector
(see e.g.\ \cite{GolPooSaf-book,Cor-book} and references therein).
It lies in the plane of motion,
which is orthogonal to the angular momentum vector,
and points in the direction given by 
the periapsis in the case of inverse-square motion 
and the position of closest separation in the case of two-body motion. 

In contrast to the conservation of energy and angular momentum in the Kepler problem,
which come directly from Noether's theorem
using the manifest time-translation and rotation symmetries of Newtonian space, 
the LRL vector does not arise from any kinematic symmetry principle. 
Nevertheless, 
Noether's theorem in reverse (e.g.\ \cite{Olv-book,Blu.Anc-book})
states that every conserved quantity
is generated by some infinitesimal symmetry transformation
that leaves invariant the action principle.
Since the LRL vector comprises three independent conserved quantities,
its symmetry generator consists of three independent infinitesimal transformations,
which were first worked out in \Ref{Lev}.
They turn out to involve the velocity variable,
unlike the infinitesimal time-translation and rotation transformations
which involve only the time and position variables.
In particular, the LRL infinitesimal symmetry transformations are primary examples of
what are called \emph{dynamical symmetries},
in contradistinction to \emph{point symmetries}
which are typified by infinitesimal time-translations and rotations. 

When the LRL symmetries are normalized to have the same physical dimension as the rotations,
these 6 point symmetries are known to form a 6-dimensional Lie algebra
whose structure turns out to be $\alg{so}(4)$ when the energy is negative,
and $\alg{so}(3,1)$ when the energy is positive \cite{Bar}. 
These Lie algebras are isomorphic to the algebra of infinitesimal isometries of,
respectively, a 3-sphere $S^3$ in $\Rnum^4$ 
and a 3-hyperboloid $H^3$ in Minkowski space  $\Rnum^{3+1}$. 

One explanation for the appearance of four-dimensional spaces is provided by
Moser's observation \cite{Mos} that, in the negative energy case,
the Kepler problem is equivalent to non-affine geodesic motion on $S^3$
under stereographic projection. 
A kinematical description of the projection is given by \cite{Foc,Bar}
identifying the stereographic coordinates in an auxiliary $\Rnum^4$ 
as the components of the Kepler velocity variable. 
The rotations in the auxiliary space constitute a $SO(4)$ transformation group
that combines the physical $SO(3)$ rotations and the LRL symmetry transformations.
A similar equivalence \cite{Bel} holds in the positive energy case
by considering non-affine geodesic motion on $H^3$.

One remaining question is how does the group generated by 
the LRL infinitesimal symmetry transformations act
just on the time, position, and velocity variables in the Kepler problem. 
It is hard to find an explicit description of this group in the literature
(although a description in terms of the coordinates in the auxiliary 4-dimensional space
is known \cite{Rod}). 

The first main result of the present work will be to derive
the explicit form of the LRL symmetry transformation group
in terms of the kinematic physical variables.

Another remaining question, which is not commonly discussed in the literature,
arises from the magnitude of LRL vector
which is given by a scalar expression in terms of
the energy and magnitude of the angular momentum.
On the one hand, 
this choice of normalization of the LRL vector is essential for
the algebra generated by the LRL symmetries and rotations
to be $\alg{so}(4)$ when energy is negative and $\alg{so}(3,1)$ when energy is positive. 
Yet on the other hand, it is a free choice, which raises the question of whether
a simpler structure is possible for the symmetry algebra
if some other choice of normalization is made.
A natural choice is to normalize it to be a unit vector,
retaining only the physically relevant direction information contained in the LRL vector.
This will be called the \emph{LRL direction vector}. 

As a second main result,
the components of the LRL direction vector will be shown to
generate three infinitesimal symmetry transformations
that commute with themselves. 
This constitutes a symmetry subalgebra $\Rnum^3$
which is abelian and belongs to the larger 6-dimensional symmetry algebra
$\alg{so}(3)\rtimes \Rnum^3$ 
given by a semi-direct sum with rotations $\alg{so}(3)$. 
The explicit form of the corresponding symmetry transformation group,
which is $SO(3) \rtimes \Rnum^3$ 
acting on the time, position, and velocity variables in the Kepler problem,
will be presented.

The rest of the paper is organized as follows. 
In section~\ref{sec:generators},
the generators of the infinitesimal symmetry transformations
will be reviewed using a modern formulation. 
The symmetry algebras respectively containing 
the LRL vector and the LRL direction vector
will be summarized in section~\ref{sec:PBs} in terms of Poisson brackets. 
In addition, the action of the infinitesimal symmetries on the constants of motion
will be derived. 
In section~\ref{sec:directionalLRL.group},
the explicit symmetry transformation group arising from the LRL direction vector
will be worked out on the kinematical variables.
In section~\ref{sec:LRL.group},
the derivation is extended to obtain the explicit form of the LRL symmetry transformation group on the kinematic variables.
Some concluding remarks will be given in section~\ref{sec:conclude}.

\section{Infinitesimal symmetry generators}\label{sec:generators}

The equations of motion of the Kepler problem are given by
\begin{equation}\label{eom}
  \ddot{\posvec} = -\k|\posvec|^{-2}\unitposvec
\end{equation}
where $\k$ is a positive constant and $\hat r = |\posvec|^{-1}\posvec$ is a unit vector. 
For particle motion in Newtonian gravity,
$\posvec$ is the position of the particle with respect to the central mass,
while $\k=GM$ where $M$ is the central mass and $G$ is the gravitational constant.
For two body motion under Newtonian gravitational attraction,
$\posvec$ is the relative position of the bodies with respect to their center of mass,
and $\k=G\mu$ where $\mu$ is the reduced mass of the two bodies. 

The constants of motion consist of
energy
\begin{equation}\label{ener}
  E = \tfrac{1}{2} |\dot{\posvec}|^2 - \k |\posvec|^{-1}, 
\end{equation}
angular momentum
\begin{equation}\label{angular.mom}
\L  = \posvec \times \dot{\posvec}, 
\end{equation}
and the LRL vector
\begin{equation}\label{lrl}
  \A = \dot{\posvec}\times \L - \k \unitposvec . 
\end{equation}
Here these quantities have been normalized by dividing out
the particle mass in the case of inverse-square motion 
and the reduced mass in the case of two-body motion.

The LRL vector is orthogonal to the angular momentum,
$\A\cdot\L =0$,
and has magnitude
$|\A| = \sqrt{\k^2 + 2 E |\L|^2}$. 
For circular orbits, $\A$ vanishes,
whereas for elliptical, parabolic,  and hyperbolic orbits,
$\A$ is non-trivial and points in the direction of the periapsis.
Its magnitude is  proportional to the eccentricity of the orbit, $e = |\A|/\kappa$. 
Therefore, among the three components of $\A$,
only the angle of $\A$ with respect to a fixed axis 
in the plane of motion for non-circular orbits 
represents an independent constant of motion.

Thus, $E$, $\L$, $\Theta$ provide 5 independent constants of motion.
This is the maximal number for any type of motion in three dimensions.
There is an additional conserved quantity,
which involves $t$ and thus is an integral of motion rather than a constant of motion,
that can be defined as the time $T$ at which the periapsis point is reached
in a non-circular orbit. 
An explicit formula for $T$, along with discussion of its physical significance,
is given in \Ref{Anc.Mea.Pas}.

Noether's theorem gives a \emph{one-to-one correspondence}
between conserved quantities 
and infinitesimal dynamical symmetries of an action principle. 
For the Kepler problem \eqref{eom}, the action principle is
$S = \int_{t_1}^{t_2} \Lagr(\posvec,\dot{\posvec})\,dt$
where
$\Lagr(\posvec,\dot{\posvec}) = \tfrac{1}{2} |\dot{\posvec}|^2 + \k |\posvec|^{-1}$
is the Lagrangian,
whose variational derivative yields the equations of motion, 
$\delta\Lagr/\delta \posvec = - \ddot{\posvec} - \k|\posvec|^{-2}\unitposvec$. 
Recall that adding a total time derivative to $\Lagr$,
which just changes $S$ by the addition of an endpoint term,
does not affect the equations of motion.
In the Lagrangian setting,
an \emph{infinitesimal dynamical symmetry} is a generator
acting on the dynamical variable $\posvec$,
\begin{equation}\label{X}
  \X = \P(t,\posvec,\dot{\posvec})\cdot\partial_{\posvec},
\end{equation}
such that the action principle is invariant up to an endpoint term.
Specifically, the condition of invariance is given by 
$\pr\X(S) = W|_{t_1}^{t_2}$
for some function $W(t,\posvec,\dot{\posvec})$,
where
$\pr\X = \P\cdot\partial_{\posvec} + \dot{\P}\cdot\partial_{\dot{\posvec}}$
denotes the prolongation of $\X$, acting on the Lagrangian variables
$\posvec$ and $\dot{\posvec}$.
Infinitesimal symmetry invariance can be usefully restated entirely
in terms of the Lagrangian as
\begin{equation}\label{inv.cond}
  \pr\X(\Lagr) = \parder{\Lagr}{\posvec} \cdot\P + \parder{\Lagr}{\dot{\posvec}} \cdot\dot{\P}
  =  \totder{W}{t} . 
\end{equation}

\begin{prop}\label{prop:noether}
A function $C(t,\posvec,\dot{\posvec})$ is a conserved quantity for the Kepler problem,
namely $\dot{C} = 0$ for all solutions of the equations of motion  \eqref{eom},
if and only if
the generator $\X =\P(t,\posvec,\dot{\posvec})\cdot\partial_{\posvec}$
is an infinitesimal dynamical symmetry of the Lagrangian $\Lagr$, 
where
$\P = \dfrac{\partial C}{\partial \dot{\posvec}}$. 
This states the Noether correspondence in an explicit form. 
\end{prop}

Any infinitesimal dynamical symmetry gives rise to
a (one-parameter) transformation group of dynamical symmetries 
which geometrically represents the flow generated by the vector field $\pr\X$.
This can be understood in two different ways.

The first way is to work in the space of Lagrangian variables $(\posvec,\dot{\posvec})$,
or equivalently, in phase space (Hamiltonian) variables 
$(\posvec,\velvec)$,
where $\velvec = \dot{\posvec}$ is the velocity vector.
Let $\Dt$ denote time derivatives evaluated through use of the equations of motion \eqref{eom},
in particular
$\Dt \posvec =\velvec$ and $\Dt \velvec = -\k|\posvec|^{-2}\unitposvec$,
whereby it acts as an operator on phase space,
\begin{equation}\label{Dt}
 \Dt = \velvec\cdot\partial_{\posvec} -\k|\posvec|^{-2}\unitposvec \cdot\partial_{\velvec} . 
\end{equation}
Then the resulting symmetry transformation group
$(\posvec,\velvec) \to (\posvec,\velvec)^*$ 
is given by
\begin{equation}\label{X.phasespace}
  (\posvec,\velvec)^* = \exp(\epsilon\,\hat\X) (\posvec,\velvec),
  \quad
  \hat\X = \P\cdot\partial_{\posvec} + \Dt\P\cdot\partial_{\velvec}
\end{equation}
with $\varepsilon$ being the group parameter,
where $\varepsilon=0$ yields the identity transformation.

A second way to define an equivalent transformation group is
by having the transformations act directly on solutions
$(\posvec(t),\velvec(t))$ of the equations of motion \eqref{eom}:
\begin{equation}\label{X.solns}
  (\posvec(t),\velvec(t))^* = \exp(\varepsilon\,\pr\X) (\posvec(t),\velvec(t)) 
\end{equation}
Both transformation groups \eqref{X.phasespace} and \eqref{X.solns}
represent the one-parameter dynamical symmetry group generated by $\X$.
Note that $t$ is kept fixed in these transformations. 

A special case of dynamical symmetry groups comprises point symmetry groups
(see e.g.\ \Ref{Blu.Anc-book,Olv-book})
which are commonly expressed by point transformations 
\begin{equation}
  \posvec\to \posvec{}^* = \posvec + \varepsilon \vec{\xi} + O(\varepsilon^2),
  \quad
  \velvec\to \velvec{}^* = \velvec + \varepsilon \dot{\vec{\xi}} + O(\varepsilon^2),
  \quad
  t\to t{}^* = t + \varepsilon \tau + O(\varepsilon^2)
\end{equation}
where $\vec{\xi}$ and $\tau$ are functions only of $t$ and $\posvec$. 
Their infinitesimal generator is given by
$\Y_\text{point} = \tau \partial_{t} + \vec{\xi}\cdot\partial_{\posvec} + \dot{\vec{\xi}}\cdot\partial_{\velvec}$, 
corresponding to the transformation group
$(t,\posvec,\velvec)^* = \exp(\varepsilon\,\Y_\text{point}) (t,\posvec,\velvec)$.
These symmetries act in the coordinate space $(t,\posvec,\velvec)$,
unlike dynamical symmetries which act in phase space $(\posvec,\velvec)$
or on solutions of the equations of motion. 
Nevertheless, point symmetries have an equivalent formulation
in terms of a generator \eqref{X} acting in phase space, 
which arises from how they geometrically map
solution curves $(\posvec(t),\velvec(t))$ into themselves 
in the coordinate space $(t,\posvec,\velvec)$. 
This mapping, in which $t$ is kept fixed, is given by \cite{Blu.Anc-book}
\begin{equation}\label{point.transformation}
  \posvec{}^*(t) = \posvec(t) + \varepsilon \big(\vec{\xi}(t,\posvec(t)) - \tau(t,\posvec(t))\,  \velvec(t)\big) + O(\varepsilon^2)
\end{equation}
along with $\velvec{}^*(t) = \frac{d}{dt}\posvec{}^*(t)$.
The resulting infinitesimal symmetry generator in phase space is 
\begin{equation}\label{hatX.pointsymm}
  \hat\X_\text{point} = \P_\text{point}\cdot\partial_{\posvec} + \Dt\P_\text{point}\cdot\partial_{\velvec},
  \quad
   \P_\text{point} = \vec{\xi} - \tau \velvec ,
\end{equation}
which is related to the previous generator 
$\Y_\text{point} = \hat\X_\text{point} + \tau D_t$,
where
\begin{equation}\label{D_t}
  D_t = \partial_t + \velvec\cdot\partial_{\posvec} + \Dt\velvec\cdot \partial_{\velvec} . 
\end{equation}
In particular, when acting on solutions $(\posvec(t),\velvec(t))$,
the vector field $\tau D_t$ represents a trivial infinitesimal symmetry
that leaves every solution unchanged. 

An important corollary observation is that any infinitesimal dynamical symmetry \eqref{X}
can be extended to act in the coordinate space $(t,\posvec,\velvec)$
by means of the generator 
\begin{equation}\label{dyn.symm.Y}
  \Y = \hat\X + \tau D_t
  = \tau\partial_{t} +\big(\P + \tau\velvec\big)\cdot\partial_{\posvec} + \big(\Dt\P - \tau |\posvec|^{-2}\unitposvec\big)\cdot\partial_{\velvec}
\end{equation}
where $\tau$ is a completely arbitrary function of $t$, $\posvec$, $\velvec$.
The flow of this vector field \eqref{dyn.symm.Y} 
generates the transformation group
\begin{equation}\label{dyn.symm.transformation}
(t,\posvec,\velvec)\to  (t,\posvec,\velvec)^* = \exp(\varepsilon\,\Y) (t,\posvec,\velvec)
\end{equation}
with parameter $\varepsilon$.
Here $\tau$ constitutes a \emph{gauge freedom},
since both $\pr\X$ and $\Y$ act in an identical way on solutions $(\posvec(t),\velvec(t))$.
Existence of this gauge freedom has not been widely recognized in the literature.
However, it will turn out to play a key role in finding the explicit form of
dynamical symmetry transformations generated from the LRL vector. 
(Index notation will be adopted hereafter, with raising and lowering via the Euclidean metric $\delta_{ij}$.)

\subsection{Infinitesimal symmetry generators from the constants of motion}

Proposition~\ref{prop:noether} gives a one-to-one Noether correspondence 
between constants of motion
and time-independent infinitesimal dynamical symmetries
for the Kepler problem.

Energy \eqref{ener} and angular momentum \eqref{angular.mom} respectively give
\begin{equation}\label{P.E}
  P^i[E] = \parder{E}{\dot{r}^i} = \dot{r}^i,
  \qquad
  \X_{E} = \dot{r}^i \partial_{r^i}
\end{equation}
and
\begin{equation}\label{P.L}I
  P^i[L^j] = \parder{L^j}{\dot{r}^i} = \epsilon_{ijk} r^k,
  \qquad
  \X_{L^j} = \epsilon_{ijk} r^k \partial_{r^i}
\end{equation}
which have the form of point symmetries \eqref{hatX.pointsymm},
namely $P^i$ is strictly linear in $\dot{r}^i=v^i$.

The generator $\X_{E}$ produces the infinitesimal symmetry 
$r^i\to   r^i{}^* = r^i +\varepsilon \dot{r}^i + O(\varepsilon^2)$
with group parameter $\varepsilon$,
yielding the symmetry transformation group on solutions 
\begin{equation}
  r^i(t){}^* = \exp(\varepsilon \hat\X_{E}) r^i(t) = \exp\Big(\varepsilon \frac{d}{dt}\Big) r^i(t)
  = r^i(t+\epsilon)
\end{equation}
and  $v^i(t){}^* =\frac{d}{dt}r^i(t){}^* = v^i(t+\epsilon)$, 
where $\hat\X_{E} = \dot{r}^i(t) \partial_{r^i} + \dot{v}^i(t) \partial_{v^i}$.
This is a time-translation symmetry group,
which is equivalent to the point transformation
\begin{equation}
  t \to t^* = t -\varepsilon,
  \quad
  r^i \to r^i{}^* = r^i,
  \quad
  v^i \to v^i{}^* = v^i
\end{equation}
acting in the coordinate space $(t,r^i,v^i)$.
Note that the minus sign in the transformation of $t$
comes from the general relation between
a point transformation \eqref{point.transformation}
and its generator \eqref{hatX.pointsymm} on phase space, 
which shows that $\tau = -1$ and $\xi^i =0$. 

The angular momentum generator $\X_{L^j}$ produces 
$r^i\to   r^i{}^* = r^i +   \epsilon^{ijk} \varepsilon_j r_k   + O(\varepsilon^2)$
with group parameter $\varepsilon^j$ (which is a vector).
This is manifestly an infinitesimal rotation acting on $r^i$.
Note that its generator \eqref{hatX.pointsymm}
has $\tau=0$ and $\xi^i = \epsilon^{ijk}r_k$.  
The corresponding symmetry transformations on solutions 
are given by the rotation group 
\begin{equation}\label{rotation.group}
  r^i(t){}^* = \exp(\varepsilon^j \hat\X_{L^j}) r^i(t) = R^i{}_j(\varepsilon) r^j(t)
\end{equation}
and $v^i(t){}^* =\frac{d}{dt}r^i(t){}^* = R^i{}_j(\varepsilon) v^j(t)$,
where $\hat\X_{L} = \epsilon_{ijk} r^k(t) \partial_{r^i} + \epsilon_{ijk} v^k(t) \partial_{v^i}$,
with
\begin{equation}\label{rotation.op}
  R^i{}_j(\varepsilon) = 
  \cos(|\varepsilon|) \delta^i{}_j +(1-\cos(|\varepsilon|)) \hat\varepsilon^i \hat\varepsilon^j +  \sin(|\varepsilon|) \hat\varepsilon_k \epsilon^{ki}{}_j,
  \quad
  |\varepsilon| = \varepsilon^j \varepsilon_j,
  \quad
    \hat\varepsilon^i = |\varepsilon|^{-1} \varepsilon^i
\end{equation}
being the familiar rotation operator. 
In particular, the rotation angle is $|\varepsilon|$
and the rotation axis is $\hat\varepsilon^i$,
determined by the group parameter $\varepsilon^i$.
These symmetries acts as a point transformation in the coordinate space $(t,r^i,v^i)$:
\begin{equation}
  t \to t^* = t, 
  \quad
  r^i \to r^i{}^* = R^i{}_j(\varepsilon) r^j, 
  \quad
  v^i \to v^i{}^* = R^i{}_j(\varepsilon) v^j .
\end{equation}

Finally, the LRL vector \eqref{lrl} yields
\begin{equation}\label{P.A}
  P^i[A^j] = \parder{A^j}{\dot{r}^i} = 2\dot{r}^i r^j - r^i \dot{r}^j - \delta^{ij} \dot{r}^k r_k , 
  \qquad
  \X_{A^j} = (2\dot{r}^i r^j - r^i \dot{r}^j -\delta^{ij} \dot{r}^k r_k)\partial_{r^i} .
\end{equation}
In contrast to the previous generators \eqref{P.E} and \eqref{P.L}, 
this expression is not strictly linear in $\dot{r}^i=v^i$, 
namely
$\displaystyle \parder{P^i[A^j]}{\dot{r}^k}
= 2\delta_{ik} r_j - \delta_{jk} r_i -\delta_{ij} r_k$
is not proportional to $\delta_{ik}$.
As a consequence, it represents the generator of a genuine dynamical symmetry
rather than a point symmetry. 
The resulting dynamical symmetry transformations on solutions $(r^i(t),v^i(t))$
involve the prolongation of this generator \eqref{P.A} to phase space.
Its components consist of $P^i[A^j]$ and $\Dt P^i[A^j]$,
where $\Dt r^k =v^k$ and $\Dt v^k  = -\k r^{-1}\hat r^k$,
with $\r = \sqrt{r^k r_k} = |\posvec|$. 
This yields
\begin{equation}\label{lrl.hatX}
\hat\X_{A^j}  =
\big( 2 v^i r^j - r^i v^j -\delta^{ij} v^k r_k \big)\partial_{r^i}
+ \big(   v^i v^j  -\k \r^{-3} r^i r^j - \delta^{ij} (2E + \k \r^{-1})  \big)\partial_{v^i} .
\end{equation}

Thus, 
the dynamical symmetry transformations arising from the LRL vector
are given by 
\begin{equation}\label{lrl.solns}
  r^i(t){}^* = \exp(\varepsilon^j \hat\X_{A^j}) r^i(t), 
  \quad
  v^i(t){}^* = \exp(\varepsilon^j \hat\X_{A^j}) v^i(t).
\end{equation}
The explicit form of these transformations
will be obtained in section~\ref{sec:LRL.group}.

As pointed out in section~\ref{sec:intro},
the underlying constants of motion $E$, $L^i$, $A^i$ are not all independent,
since the magnitude of $A^i$ is a function of $E$ and $|L| = \sqrt{L^i L_i}$.
The physically relevant content of $A^i$ is its direction,
which is given by the unit vector
\begin{equation}\label{lrl.direction}
  \Theta^i : = |A|^{-1} A^i 
\end{equation}
where $|A|= \sqrt{A^j A_j} =\sqrt{\k^2 + 2E |L|^2}$. 
This direction vector \eqref{lrl.direction} is a constant of motion,
and hence it gives rise to an infinitesimal symmetry just like $A^i$ does.
Proposition~\ref{prop:noether} yields
\begin{equation}\label{P.Theta}
  P^i[\Theta^j] = \parder{\Theta^j}{\dot{r}^i}
  = |A|^{-1} (2\dot{r}^i r^j - r^i \dot{r}^j - \delta^{ij} \dot{r}^k r_k)
  -|A|^{-3} (2E \epsilon^{ikl} L_k r_l +\tfrac{1}{2} |L|^2 \dot{r}^i) A^j
\end{equation}
Hence, the infinitesimal symmetry generator is given by 
\begin{equation}\label{X.Theta}  
  \X_{\Theta^j} = \big( |A|^{-1} (2v^i r^j - r^i v^j - \delta^{ij} v^k r_k)
  - |A|^{-3} (2 E \epsilon^{ikl} L_k r_l  + |L|^2 v^i)A^j \big)  \partial_{r^i} . 
\end{equation}
This represents a genuine dynamical symmetry,
similarly to the previous generator \eqref{P.A}.

To prolong the generator \eqref{X.Theta} to phase space,
its component $\Dt P^i[\Theta^j]$ must be computed.
Since $L_k$, $E$, $A_k$ are constants of motion,
they satisfy $\Dt L_k = \Dt E = \Dt A_k =0$.
Hence, 
\begin{equation}
\Dt P^i[\Theta^j] = 
|A|^{-1} (   v^i v^j  -\k \r^{-3} r^i r^j - \delta^{ij} (2E + \k \r^{-1})  )
- |A|^{-3}(2 E \epsilon^{ikl} L_k v_l -\kappa |L|^2 \r^{-3} r^i ) A^j . 
\end{equation} 
Thus, the prolonged generator in phase space is 
\begin{equation}
  \begin{aligned}\label{directionlrl.hatX}
  \hat\X_{\Theta^j} & =
  \big( |A|^{-1} ( 2v^i r^j - r^i v^j - \delta^{ij} v^k r_k )
  -|A|^{-3} (2E \epsilon^{ikl} L_k r_l +|L|^2 v^i) A^j \big)  \partial_{r^i} 
  \\&\qquad
  + \big(  |A|^{-1} (   v^i v^j  -\k \r^{-3} r^i r^j - \delta^{ij} (2E + \k \r^{-1})  )
  - |A|^{-3}(2E \epsilon^{ikl} L_k v_l  -\k |L|^2 \r^{-3} r^i ) A^j 
  \big)\partial_{v^i}, 
  \end{aligned}
\end{equation}
and the resulting dynamical symmetry transformations are given by 
\begin{equation}\label{directionlrl.solns}
  r^i(t){}^* = \exp(\varepsilon^j \hat\X_{\Theta^j}) r^i(t), 
  \quad
  v^i(t){}^* = \exp(\varepsilon^j \hat\X_{\Theta^j}) v^i(t).
\end{equation}
Their explicit form will be obtained in section~\ref{sec:directionalLRL.group}.

\section{Poisson brackets}\label{sec:PBs}

The Kepler problem \eqref{eom} has the Hamiltonian formulation 
$\displaystyle \dot{\posvec} = \parder{H(\posvec,\velvec)}{\velvec}$, 
$\displaystyle \dot{\velvec} = -\parder{H(\posvec,\velvec)}{\posvec}$
where
$H(\posvec,\velvec) =   \tfrac{1}{2} |\velvec|^2 - \k |\posvec|^{-1}$
is the energy \eqref{ener}.
The Poisson bracket is given by 
\begin{equation}
  \{ F,G \} = \parder{F}{\posvec}\cdot \parder{G}{\velvec} - \parder{G}{\posvec}\cdot \parder{F}{\velvec}
\end{equation}
for any functions $F(\posvec,\velvec)$ and $G(\posvec,\velvec)$ on phase space.
When these functions are constants of motion, note that
$\displaystyle\parder{F}{\velvec} = \vec{P}[F]$
and $\displaystyle\parder{G}{\velvec} = \vec{P}[G]$
are given by the corresponding infinitesimal symmetry generators
through Noether's theorem as stated in Proposition~\ref{prop:noether}. 

As is well known,
the constants of motion --- energy, angular momentum, LRL vector ---
form a Lie algebra under the Poisson bracket.
To compute the brackets,
it is simplest to work in index notation (cf Section~\ref{sec:generators})
and use expressions \eqref{P.E}, \eqref{P.L}, \eqref{P.A}. 
This yields
\begin{gather}
  \{ E, L^i \} 
  =0 , 
\label{E.L.bracket}
\quad
\{ L^i, L^j \}
   = \epsilon^{ijk} L_k , 
\\
  \{ E, A^i \}
     =0 , 
\quad
\{ A^i, L^j \}
   = \epsilon^{ijk} A_k , 
\quad
\{ A^i, A^j \}
   = -2 E \epsilon^{ijk} L_k .
\label{A.A.bracket}
\end{gather}

Thus, $E$, $L^i$, and $A^i$ generate a closed algebra of Poisson brackets.
Its structure is well known to take the simplest form if $A^i$ is replaced by
\begin{equation}\label{M}
M^i = \sqrt{2|E|}^{-1} A^i
\end{equation}
assuming $E\neq0$. 
Notice that $M^i$ is a constant of motion and has the same physical dimensions as $L^i$.
This yields the 7-dimensional algebra
\begin{subequations}\label{lrl.bracket.alg}
\begin{gather}
  \{ E,L^i \} =0,
  \quad
  \{ E,M^i \} =0,
  \quad
  \{ L^i, M^j \}  = \epsilon^{ij}{}_k M^k ,
  \\
  \{ L^i, L^j \}  = \epsilon^{ij}{}_k L^k,
  \quad
  \{ M^i, M^j \}  = -\sgn(E)\,\epsilon^{ij}{}_k L^k , 
\end{gather}
\end{subequations}
which has the structure
$\Rnum\oplus \alg{so}(4)$ when $E<0$,
and $\Rnum\oplus \alg{so}(3,1)$ when $E>0$. 

Perhaps unsurprisingly,
since $M^i$ and $L^i$ have the same physical dimensions,
the rotation algebras $\alg{so}(4)$ and $\alg{so}(3,1)$ 
that they generate cannot be factored into simpler subalgebras.
So it is natural to expect that
if $M^i$ is replaced by the dimensionless LRL direction vector \eqref{lrl.direction},
then the Lie algebra generated by $\Theta^i$ and $L^i$
should have a simpler structure.

The brackets involving $\Theta^i$ can be computed
directly from the previous brackets \eqref{E.L.bracket}--\eqref{A.A.bracket}
by using the relation $A^i =|A|\, \Theta^i$ together with the Leibniz rule.
In particular, for any function $F(r^k,v^k)$ on phase space,
note that its bracket with $\Theta^i$ can be obtained from 
$\{ F, A^i \}
= |A|\ \{ F, \Theta^i \} + \Theta^i\ \{ F, |A| \}$,
where $\{ F, |A| \} =  |A|^{-1}A_j\ \{ F, A^j\}$. 
This gives
\begin{align}
  \{ E, |A| \}  
  = 0 , 
\quad
   \{ L^i, |A| \}  
   =0 , 
\quad
   \{ A^i, |A| \}  
   = -2 |A|^{-1} E \epsilon^{ijk} A_j L_k ; 
\end{align}
and then secondly, 
$\{ F, \Theta^i \} = |A|^{-1}( \{ F, A^i \} - \Theta^i\ \{ F, |A| \} )$
yields 
\begin{gather}
\{ E, \Theta^i \} =0 , 
\quad
 \{ L^j, \Theta^i \}
=\epsilon^{jik} \Theta_k , 
\\
\{ |A|,\Theta^ i \} = 2 E \epsilon^{ijk} \Theta_j L_k , 
\quad
\{ A^j, \Theta^i \}   = 2 |A|^{-1}E (  \epsilon^{jlk} \Theta_l \Theta^i - \epsilon^{jik} ) L_k , 
\quad
  \{ \Theta^j, \Theta^i \}  
  = 0 , 
 \label{Theta.Theta.bracket}
\end{gather}
where the last line has used the vector identity
$\epsilon^{jlk} F^i - \epsilon^{ilk} F^j = \epsilon^{jik} F^l - \epsilon^{jil} F^k$,
as well as the relations $\Theta^j L_j =0$ and $\Theta^j \Theta_j =1$.
Therefore,
only the bracket $\{ L^i, \Theta^j \}$ among $E$, $L^i$, $\Theta^i$ is non-zero.
It shows that $L^j$ acts as an infinitesimal rotation on the direction
vector $\Theta^i$. 

\begin{thm}\label{thm:lrldirection.alg}
The constants of motion $E$, $L^i$, and $\Theta^i$
generate a 7-dimensional Lie algebra under the Poisson bracket,
which has the structure
$\Rnum\oplus \alg{so}(3)\rtimes \Rnum^3$. 
This algebra is isomorphic to the Lie algebra of the infinitesimal symmetries
associated to these constants of motion. 
\end{thm}

\subsection{Symmetry action on constants of motion}

The set of all constants of motion of the  Kepler problem forms a vector space,
and any infinitesimal symmetry $\hat\X$ maps this vector space into itself. 
The mapping is explicitly given by the Poisson bracket
$\pr\X_{C} f = \{f,C\}$, 
as shown by a direct computation. 
Therefore, 
the only non-zero actions of the infinitesimal symmetries
$\hat\X_{E}$, $\hat\X_{L^j}$, $\hat\X_{A^j}$, $\hat\X_{\Theta^j}$ 
on the constants of motion $E$, $L^i$, $A^i$, and $\Theta^i$
are given by
\begin{subequations}
\begin{gather}
 \pr\hat\X_{L^j}(L^i) = \epsilon^{ij}{}_k L^k , 
\quad
 \pr\hat\X_{A^j}(A^i) = - 2 E \epsilon^{ij}{}_k L^k , 
\quad
 \pr\hat\X_{L^j}(A^i) = -\pr\hat\X_{A^i}(L^j)  = \epsilon^{ij}{}_k A^k , 
  \\
\pr\hat\X_{L^j}(\Theta^i) = - \pr\hat\X_{\Theta^i}(L^j)  = \epsilon^{ij}{}_k \Theta^k , 
\
 \pr\hat\X_{A^j}(\Theta^i) = -\pr\hat\X_{\Theta^i}(A^j) 
 = 2 |A|^{-1}E  \epsilon^{jlk}(\delta_l^i - \Theta_l \Theta^i) L_k . 
\end{gather}
\end{subequations}

Similarly, from the brackets \eqref{lrl.bracket.alg}, 
observe that the 7-dimensional vector space spanned by $E$, $L^i$, $M^i$
carries a homomorphic representation of the LRL symmetry algebra.
The same statement holds for the symmetry algebra in Theorem~\ref{thm:lrldirection.alg}
on the span of $E$, $L^i$, $\Theta^i$.

\section{Symmetry group generated by the directional LRL vector}\label{sec:directionalLRL.group}

The dynamical symmetry transformations \eqref{directionlrl.solns}
arising from the LRL direction vector \eqref{lrl.direction}
are most readily expressed in an explicit form
by working in the coordinate space $(t,\posvec,\velvec)$
where the generator of the infinitesimal symmetry 
possesses gauge freedom which can be exploited to simplify the transformations.
This generator has the form 
\begin{equation}\label{directionlrl.Y}
  \Y_{\Theta^j}   = \tau^j D_{t} + \hat\X_{\Theta^j} 
\end{equation}
where $\hat\X_{\Theta^j}$ is the generator \eqref{directionlrl.hatX} in phase space, 
and $\tau^j$ is an arbitrary function of $t$, $\posvec$, $\velvec$.
Now $\tau^j$ can be chosen so that one component of this generator \eqref{directionlrl.Y}
is zero, 
which will consequently be invariant under the infinitesimal symmetry \eqref{directionlrl.Y}.

It will be useful to have $\r$ be invariant,
$\Y_{\Theta^j}(\r) =0$. 
This condition is easily computed via 
\begin{equation}
  \begin{aligned}
    \r\Y_{\Theta^j}(\r) = r_i \Y_{\Theta^j}(r^i) 
    & = \tau^j r_i v^i +|A|^{-2} (|L|^2 -\k \r) \epsilon^{jkl}\Theta_k L_l
  \end{aligned}
\end{equation}
which determines 
\begin{equation}\label{gauge.directionlrl}
  \tau^j  =   |A|^{-2} (\k \r - |L|^2) (r_i v^i)^{-1} \epsilon^{jkl} \Theta_k L_l. 
\end{equation}

The components of the gauge-fixed generator 
represent a vector field in the coordinate space $(t,r^i,v^i)$.
Its flow thereby defines the dynamical symmetry transformations
\begin{equation}\label{directionlrl.flow}
(t,r^i,v^i)\to   (t,r^i,v^i)^* = \exp(\varepsilon^j \Y_{\Theta^j}) (t,r^i,v^i)
\end{equation}
acting in this space,
where $\varepsilon^j$ is the parameter of the flow.
The components of this flow \eqref{directionlrl.flow} satisfy
the system of 7 coupled differential equations
\begin{equation}\label{sys.directionlrl.flow}
  \totder{t^*(\varepsilon)}{\varepsilon_j} 
  = \Y_{\Theta^j}^*[\varepsilon]  (t^*(\varepsilon)) ,
  \quad
  \totder{r^i{}^*(\varepsilon)}{\varepsilon_j}
  = \Y_{\Theta^j}^*[\varepsilon](r^i{}^*(\varepsilon)) ,
    \quad
  \totder{v^i{}^*(\varepsilon)}{\varepsilon_j}
  = \Y_{\Theta^j}^*[\varepsilon](v^i{}^*(\varepsilon)) 
\end{equation}
where $\Y_{\Theta^j}^*[\varepsilon] = \Y_{\Theta^j}\big|_{t=(t^*(\varepsilon),r^i= r^i{}^*(\varepsilon),v^i=v^i{}^*(\varepsilon)}$
denotes the generator evaluated on the flow.
In addition, the flow satisfies the condition that $\varepsilon^j=0$
gives the identity transformation, 
\begin{equation}\label{sys.iv}
  t^*(0)  = t,
  \quad
  r^i{}^*(0) = r^i,
  \quad
  v^i{}^*(0) = v^i . 
\end{equation}
The goal now will be to integrate this system explicitly.
Hereafter, for notational simplicity,
the dependence on $\varepsilon^j$ will be suppressed
in the transformed variables and the generator.

As a starting point,
the gauge condition implies 
$\displaystyle 
  \totder{\r^*}{\varepsilon_j} 
  = \Y_{\Theta^j}^*(\r^*) =0$, 
and therefore
\begin{equation}\label{scalar.r.soln}
  \r^* = \r . 
\end{equation}
constitutes the solution of one combination of the 7 differential equations
in the system \eqref{sys.directionlrl.flow}. 
The strategy for integrating the remainder of the system involves three steps:
(1) Convert the differential equations for $r^i{}^*$ and $v^i{}^*$ 
into an equivalent simpler system with the constants of motion
$E{}^*$, $L^i{}^*$, $\Theta^i{}^*$ 
taken as the variables in the flow, 
together with the invariant $\r$.
This system is obtained from the action of the infinitesimal symmetry $\hat\X_{\Theta^i}$
on $E{}^*$, $L^i{}^*$, $\Theta^i{}^*$,
which is given by the Poisson brackets.
(2) Express $r^i{}^*$ and $v^i{}^*$ algebraically 
in terms of $E{}^*$, $L^i{}^*$, $\Theta^i{}^*$, and $\r$.
This amounts to a straightforward linear algebra computation.
(3) Directly integrate the remaining differential equation for $t{}^*$,
as its righthand side will involve only solved quantities.
Details of these steps will be given elsewhere \cite{Anc.Mah.2026a}. 

\begin{thm}\label{thm:directionLRL.group}
The dynamical symmetry transformations \eqref{directionlrl.flow},
which arise from the LRL direction vector \eqref{lrl.direction},
have the explicit form 
\begin{subequations}\label{directionLRL.transformations}
\begin{gather}
 t{}^* = t +  \T(r,|\L^*|,E) - \T(r,|\L|,E) ,
  \label{directionLRL.t}
  \\
\posvec{}^* =  |\A^*|^{-1}(  \alpha^r{}^* \vec{\Theta} + \beta^r{}^* \L^*\times\vec{\Theta} ),
  \quad
  \velvec{}^* = |\A^*|^{-1}( \alpha^v{}^* \vec{\Theta} + \beta^v{}^* \L^* \times \vec{\Theta} )
  \label{directionLRL.r.v}
\end{gather}
\end{subequations}
with parameter $\vec{\varepsilon}$, 
where
\begin{equation}\label{directionLRL.L}
  \L^* = \L + \vec{\varepsilon}\times \vec{\Theta} 
\end{equation}
and 
\begin{equation}\label{r.v.coeffs}
  \alpha^r{}^* = |\L^*|^2 - \k |\posvec|, 
  \quad
  \beta^r{}^* =   \gamma^*, 
\quad
  \alpha^v{}^* = -\k |\posvec|^{-1} \gamma^*, 
  \quad
  \beta^v{}^* = 2 E +\k |\posvec|^{-1} , 
\end{equation}
with 
\begin{equation}\label{T}
  \T(r,|\L|,E) =
  \begin{cases}
  \dfrac{1}{2E} \Big(  \gamma -\dfrac{\k}{\sqrt{2|E|}}\arctan\Big( \dfrac{\sqrt{2|E|}\gamma}{2E |\posvec| + \k}
  \Big)  \Big),
    & E<0 \\
    \dfrac{1}{2E} \Big(  \gamma  -\dfrac{\k}{\sqrt{2E}}\arctanh\Big( \dfrac{\sqrt{2E} \gamma}{2E|\posvec| + \k} \Big)  \Big),
    & E>0 \\
    \dfrac{\k |\posvec| + |\L|^2}{3 \k^2} \gamma, 
    & E=0
  \end{cases}
\end{equation}and 
\begin{equation}\label{gam}
  \gamma=    \sgn(\posvec\cdot\velvec) \sqrt{2(E|\posvec| + \k) |\posvec| - |\L|^2} . 
\end{equation}

Under these transformations, 
$E$ and $\vec{\Theta}$ are invariant, along with $|\posvec|$,
while 
\begin{equation}
  |\A^*| = \sqrt{\k^2 + 2 E|\L{\strut}^*|}
  =   \sqrt{ |\A|^2+ 2E\big( \vec{\varepsilon}\cdot(\vec{\Theta}\times\L) + |\vec{\varepsilon}|^2 - (\vec{\varepsilon}\cdot\vec{\Theta})^2 \big) } . 
\end{equation}
\end{thm}

As shown by Theorem~\ref{thm:lrldirection.alg},
the three infinitesimal symmetries $\hat\X_{\Theta^j}$ form an abelian Lie algebra $\Rnum^3$.
This implies that
the set of all dynamical symmetry transformations \eqref{directionLRL.transformations}
under composition generates a 3-dimensional abelian Lie group.
Its physical meaning is that the group transformations preserve
energy and the direction of the LRL vector,
while angular momentum is shifted by a vector in the plane orthogonal to the LRL vector.
The change in the magnitude of the LRL vector 
is proportional to the change in eccentricity.
In the case of elliptical orbits, since energy $E<0$ is unchanged,
the period $T = 2\pi/\sqrt{2|E|}$ and semi-major axis $a = \kappa/(2|E|)$
are also preserved.

\section{LRL symmetry transformation group}\label{sec:LRL.group}

The explicit form of the LRL dynamical symmetry transformations \eqref{lrl.solns}
will now be derived in a similar way to the result in Theorem~\ref{thm:directionLRL.group}.

In the coordinate space $(t,\posvec,\velvec)$,
the LRL infinitesimal symmetry generator is given by 
\begin{equation}\label{lrl.Y}
  \Y_{A^j}   = \tau^j D_{t} + \hat\X_{A^j} 
\end{equation}
where $\hat\X_{A^j}$ is the symmetry generator \eqref{lrl.hatX} in phase space, 
and $\tau^j$ is an arbitrary function of $t$, $\posvec$, $\velvec$.
This gauge freedom will be fixed by
taking $\r$ to be invariant: 
\begin{equation}
  \r\Y_{A^j}(\r)
 = r_i \Y_{A^j}(r^i) 
 = (\tau^j + r^j)v^i r_i  - v^j \r^2 
\end{equation}
which yields
\begin{equation}\label{gauge.lrl}
  \tau^j = \r^2 (v^i r_i)^{-1} v^j - r^j
  = (v^i r_i)^{-1} \epsilon^{jkl} L_k r_l . 
\end{equation}
The gauge-fixed generator \eqref{lrl.Y} 
constitutes a vector field in the coordinate space $(t,r^i,v^i)$.

The flow of this vector field determines the LRL dynamical symmetry transformations
\begin{equation}\label{lrl.flow}
(t,r^i,v^i)\to   (t,r^i,v^i)^* = \exp(\varepsilon^j \Y_{A^j}) (t,r^i,v^i)
\end{equation}
where $\varepsilon^j$ is the parameter of the flow. 
In component form,
the flow is defined by a system of 7 coupled differential equations
\begin{equation}\label{sys.lrl.flow}
  \totder{t^*(\varepsilon)}{\varepsilon_j} 
  = \Y_{A^j}^*[\varepsilon]  (t^*(\varepsilon)) ,
  \quad
  \totder{r^i{}^*(\varepsilon)}{\varepsilon_j}
  = \Y_{A^j}^*[\varepsilon](r^i{}^*(\varepsilon)) ,
    \quad
  \totder{v^i{}^*(\varepsilon)}{\varepsilon_j}
  = \Y_{A^j}^*[\varepsilon](v^i{}^*(\varepsilon)) 
\end{equation}
where $\Y_{A^j}^*[\varepsilon] = \Y_{A^j}\big|_{t=(t^*(\varepsilon),r^i= r^i{}^*(\varepsilon),v^i=v^i{}^*(\varepsilon)}$
denotes the generator evaluated on the flow.
The flow also satisfies the condition \eqref{sys.iv}
stating that $\varepsilon^j=0$ gives the identity transformation. 
For notational simplicity, hereafter
the dependence on $\varepsilon^j$ will be suppressed. 

To integrate this system,
the same three steps used in the previous section can be employed
with one change:
In the first step, the primary variables will be taken to be 
$E^*$ and $\r$, together with $M^i{}^*$ and $L^i{}^*$, 
as the latter satisfy a coupled system shown by their Poisson brackets \eqref{lrl.bracket.alg},
with $\Theta^i{}^*$ no longer being invariant in the flow. 
This leads to a more complicated transformation on $r^i{}^*$ and $v^i{}^*$
in the second step. 
Details will be given elsewhere \cite{Anc.Mah.2026a}. 

To state the main result,
introduce the matrix-rotation operator 
\begin{equation}\label{R.op}
  \underline{R}(\phi,\hat{n}) = 
  \cos(\phi)\ \underline{I}
  +(1-\cos(\phi)) \underline{I}\, \hat{n}\, \hat{n}\cdot
  +  \sin(\phi)  \underline{S}\, \hat{n}\times
\end{equation}
with
\begin{equation}
  \underline{I} =  \begin{pmatrix} 1 & 0  \\ 0 & 1 \end{pmatrix},
  \quad
  \underline{S} =  \sqrt{-\sgn(E)}\begin{pmatrix} 0 & 1  \\ -\sgn(E) & 0 \end{pmatrix}
\end{equation}
where $\phi$ is a scalar parameter and $\hat n$ is a unit-vector parameter.
For $E<0$, this operator is a twisted composition of an internal rotation $SO(2)$ 
and a spatial rotation $SO(3)$.
For $E>0$, the internal part becomes a hyperbolic rotation $SO(1,1)$.

\begin{thm}\label{thm:LRL.group}
The LRL dynamical symmetry transformations \eqref{lrl.flow}
have the explicit form
\begin{subequations}\label{LRL.transformations}
\begin{gather}
 t{}^* = t +  \int_0^{\vec{\varepsilon}}\big( |\L^*|^2 \vec{\Theta}^* + (\k |\posvec| -|\L^*|^2) \gamma^*{}^{-1} \L^*\times \vec{\Theta}^* \big)\cdot d\vec{\varepsilon} , 
  \label{LRL.t}
  \\
\posvec{}^* = |\A^*|^{-1}(  \alpha^r{}^* \vec{\Theta}^* + \beta^r{}^* \L^*\times \vec{\Theta}^* ),
  \quad
  \velvec{}^* = |\A^*|^{-1}( \alpha^v{}^* \vec{\Theta}^* + \beta^v{}^* \L^* \times \vec{\Theta}^* )
  \label{LRL.r.v}
\end{gather}
\end{subequations}
with parameter $\vec{\varepsilon}$, 
where the coefficients $\alpha^r$, $\alpha^v$, $\beta^r$, $\beta^v$
are given by expressions \eqref{r.v.coeffs}--\eqref{gam}. 

In the case $E\neq0$,
the transformed constants of motion $\L^*$ and $\vec{\Theta}^*$
are given by 
\begin{equation}\label{LRL.L.Enot0}
  \begin{pmatrix}  \L^* \\ \M^* \end{pmatrix}
  = \underline{R}(\sqrt{-2E}\, |\vec{\varepsilon}|,\hat\varepsilon)
  \begin{pmatrix}  \L \\ \M \end{pmatrix},
  \quad
  \hat\varepsilon = |\vec{\varepsilon}|^{-1} \vec{\varepsilon},
\end{equation}
in terms of $\M = |\M|\vec{\Theta}$ 
with $|\M|^2 = \sgn(E) |\L|^2 + \tfrac{1}{2}\k^2 |E|^{-1}$.
Under these transformations, 
$E$ and $|\M|^2 -\sgn(E)\, |\L|^2$, along with $|\posvec|$, are invariant,
while $|\A^*| = \sqrt{\k^2 + 2 E|\L{\strut}^*|^2}$
where
\begin{equation}
  \begin{aligned}
    |\L^*|^2  & = \cos(\sqrt{-2E}\, |\vec{\varepsilon}|)^2 |\L|^2 + \sin(\sqrt{-2E}\, |\vec{\varepsilon}|)^2 ( (\hat{\varepsilon}\cdot\L)^2 + \sgn(E) (\hat{\varepsilon}\cdot\M)^2 - \sgn(E)|\M|^2 ) \\&\qquad
    +\sqrt{-\sgn(E)} \sin(2\sqrt{-2E}) \vec{\varepsilon}\cdot(\M\times \L)
  \end{aligned}
\end{equation}

In the case $E=0$,
\begin{equation}\label{LRL.L.Eis0}
  \vec{\Theta}{}^* = \vec{\Theta}, 
  \quad
  \L^* = \L + \k\, \vec{\varepsilon}\times \vec{\Theta}
\end{equation}
are the transformed constants of motion. 
Under these transformations, 
$E$, $\vec{\Theta}$, $|\A| = \k$, 
and $|\posvec|$ are invariant. 
\end{thm}

Recall that the three infinitesimal symmetries $\hat\X_{A^j}$
do not comprise a closed algebra when $E\neq0$, 
but instead they generate a 6-dimensional Lie algebra
that contains the spatial rotations $\alg{so}(3)$
as a subalgebra generated by $\hat\X_{L^j}$
through the commutator of $[\hat\X_{A^j},\hat\X_{A^k}]$. 
Consequently,
the LRL dynamical symmetry transformations \eqref{LRL.transformations}
do not themselves form a group.
However, if the parameter $\hat{n} =\hat\varepsilon$ is taken to be fixed, 
then the set of resulting transformations generates a one-dimensional Lie group 
with parameter $\phi= \sqrt{-2E}\, |\vec{\varepsilon}|\neq0$.
For $E<0$, this group is isomorphic to the rotation group $SO(2)$
and acts on $(\L,\M)$ via
the planar rotation operator twisted with the Pauli matrix $\underline{\sigma}_1$:
\begin{equation}
  \underline{R}(\sqrt{2|E|},\hat{n}) = 
  \cos(\sqrt{2|E|})\ \underline{I}
  +(1-\cos(\sqrt{2|E|})) \underline{I}\, \hat{n}\, \hat{n}\cdot
  +  \sin(\sqrt{2|E|})  \underline{\sigma}_1\, \hat{n}\times
\end{equation}
For $E>0$, the group is isomorphic to the hyperbolic rotation group $SO(1,1)$,
with its action on $(\L,\M)$ given by
the planar boost operator twisted with the Pauli matrix $\underline{\sigma}_2$:
\begin{equation}
  \underline{R}(i\sqrt{2E},\hat{n}) = 
  \cosh(\sqrt{2E})\ \underline{I}
  +(1-\cosh(\sqrt{2E})) \underline{I}\, \hat{n}\, \hat{n}\cdot
  -\sinh(\sqrt{2E})  \underline{\sigma}_2\, \hat{n}\times
\end{equation}
In each case, the composition of all these one-dimensional groups
with $\hat{n}=\hat{\varepsilon}\in S^2$ 
will generate a 6-dimensional Lie group whose structure depends on $E$,
which is
$SO(4)$ when $E<0$
and 
$SO(3,1)$ when $E<0$.

The 6-dimensional symmetry transformation group
acts as isometries in the space of constants of motion $(\L,\M)$, 
where $\L\cdot\M=0$ and $|\L|^2 -\sgn(E)|\M|^2=\kappa^2/(2|E|)$,
with $|\L|\neq0$ and $|\M|\neq0$.
When $E<0$, the space is isomorphic to $S^2\times S^1\times S^1$,
which has the symmetry group $SO(4)$.
The space is isomorphic to $S^2\times S^1\times \Rnum$ when $E>0$,
and the symmetry group is then $SO(3,1)$.

In the case $E=0$, 
the LRL dynamical symmetry transformations \eqref{LRL.transformations} commute,
forming an abelian Lie group isomorphic to $\Rnum^3$,
which also commutes with the $SO(3)$ group generated by the rotations. 

The physical meaning of the LRL dynamical symmetry transformations \eqref{LRL.transformations}
is that they preserve energy,
while angular momentum and the LRL vector are changed in a specific way that depends on the energy. 
In the case $E<0$, the vectors $\L$ and $\M=\sqrt{2|E|}^{-1}\A$ are rotated,
preserving $|\L|^2 +|\M|^2$ and $\L\cdot\M=0$,
along with the period $T = 2\pi/\sqrt{2|E|}$ and semi-major axis $a = \kappa/(2|E|)$
of the elliptical orbit. 
Likewise in the case $E>0$,
$|\L|^2 -\frac{1}{2}|\A|^2/E$ is preserved,
as is the semi-major axis $a = \kappa/(2E)$ of the hyperbolic orbit. 
In the case $E=0$,
$\A$ is preserved and $\L$ is shifted by a vector in the plane orthogonal to $\A$.

\section{Concluding remarks}\label{sec:conclude}

In the Kepler problem, 
the dynamical symmetry transformations arising respectively 
from the LRL vector and the LRL direction vector through Noether's theorem (in reverse)
have been derived in an explicit form in terms of the kinematic variables
$t$, $\posvec$, $\velvec$.
The structure and physical meaning of the symmetry groups
generated by these respective transformations \eqref{LRL.transformations} and \eqref{directionLRL.transformations}
have been explained.

The methods and results in the present work can be extended, firstly,
to general central force dynamics. 
In particular, a generalization of the conserved LRL vector 
is known to exist for any central force
\cite{Fra,Muk,Per,Yos,Ben-Yaa,Anc.Mea.Pas} in $n>1$ dimensions.
The physically relevant part of this vector is its direction in the plane of motion,
which will give rise to an associated infinitesimal symmetry 
and thereby generate a transformation group of dynamical symmetries.
A sequel paper \cite{Anc.Mah.2026a} will study this group
as well as the group containing the generalized LRL symmetry transformations
and the temporal symmetry transformations associated to
the additional integral of motion \cite{Anc.Mea.Pas} which explicitly involves $t$. 

A further extension can be contemplated to special relativity (see e.g.\ \Ref{Fra}).
More interestingly,
as shown recently in \Ref{Anc.Faz},
an analog of the LRL vector exists for particle orbits in Schwarzschild spacetime,
where these orbits can be viewed as motion
under a relativistic, gravitational central force
\cite{MTW-book}.
Finding the transformation group of dynamical symmetries that is generated
by this analog is an interesting question which will be pursued in future work
\cite{Anc.Mah.2026b}.

\section*{Acknowledgements}

S.C.A.\ is supported by an NSERC Discovery grant.
D. Holm and C. Stoica are thanked for helpful remarks.


\begin{thebibliography}{99}

\bibitem{GolPooSaf-book}
H. Goldstein, C. Poole, J. Safko, 
{\it Classical Mechanics} (3rd ed.), (Addison Wesley) 2000.

\bibitem{Cor-book}
B. Cordani, 
{\it The Kepler Problem} (Birkhaeuser) 2003.

\bibitem{Olv-book}
P.J. Olver,
{\it Applications of Lie Groups to Differential Equations},
(Springer, New York) 1986.

\bibitem{Blu.Anc-book}
G. Bluman and S.C. Anco,
{\it Symmetry and Integration Methods for Differential Equations},
Applied Math. Sci.  154, 
(Springer, New York) 2002.

\bibitem{Lev}
J.M. L\'evy-Leblond,
Conservation laws for gauge-invariant Lagrangians in classical mechanics, 
Amer. J. Phys. 39 (1971), 502-–506. 

\bibitem{Bar}
V. Bargmann, 
Zur Theorie des Wasserstoffatoms: Bemerkungen zur gleichnamigen Arbeit von V. Fock,
Zeitschrift f\"ur Physik,  99 (7-8)  (1936), 576--582.

\bibitem{Mos}
J. Moser,
Regularization of Kepler's problem and the averaging method on a manifold,
Commun. Pure Appl. 23 (1970), 609--636. 

\bibitem{Foc}
V. Fock, 
Zur Theorie des Wasserstoffatoms,
Zeitschrift f\"ur Physik. 98 (3-4) (1935), 145--154.

\bibitem{Bel}
E.A. Belbruno,
Two-body motion under the inverse square central force and equivalent geodesic flows,
Celest. Mech. 15 (1977), 467--476.

\bibitem{Rod}
H. Rodgers,
Symmetry transformations of the classical Kepler problem, 
J. Math. Phys. 14 (1973), 1125--1129.

  
\bibitem{Anc.Mea.Pas}
S.C. Anco, T. Meadows, V. Pascuzzi,
Some new aspects of first integrals and symmetries for central force dynamics,
J. Math. Phys. 57 (2016), 062901.


\bibitem{Fra}
D.M. Fradkin,
Prog. Theor. Phys. 37 (1967), 798--812.

\bibitem{Muk}
N. Mukanda,
Dynamical symmetries and classical mechanics,
Phys. Rev. 155 (1967), 1383--1386. 

\bibitem{Per}
A. Peres,  
J. Phys. A: Math. Gen. 12 (1979), 1711--1713.

\bibitem{Yos}
  T. Yoshida,
  Two methods of generalisation of the Laplace--Runge--Lenz vector, 
  Eur. J. Phys. 8 (1987), 258--259.

\bibitem{Ben-Yaa}
  U. Ben-Ya'acov, 
Laplace--Runge--Lenz symmetry in general rotationally symmetric systems, 
J. Math. Phys. 51 (2010), 122902.
  
\bibitem{Anc.Mah.2026a}
S.C. Anco and M. Gol Bashmani Moghadam,
Conserved Laplace-Runge-Lenz quantities and symmetries in central force dynamics.
In preparation.

\bibitem{Anc.Faz}
  S.C. Anco and J. Fazio, 
Analogue of a Laplace--Runge--Lenz vector for particle orbits (timelike geodesics)
 in Schwarzschild spacetime, 
 J. Math. Phys. 64 (2023), 082501

\bibitem{MTW-book}
C.W. Misner. K.S. Thorne, J.A. Wheeler,
{\it Gravitation} (W.H. Freeman and Co.), 1973.

\bibitem{Anc.Mah.2026b}
S.C. Anco and M. Gol Bashmani Moghadam,
In preparation.  
  


\end{thebibliography}
\end{document}